\documentclass[aps, superscriptaddress, showpacs, twocolumn, longbibliography, prx]{revtex4-2}
\usepackage{units}
\usepackage{amsmath}
\usepackage{amssymb}
\usepackage{graphicx}
\usepackage{bm}
\usepackage{multirow,color,relsize,microtype,mathtools}
\usepackage[normalem]{ulem}
\usepackage{appendix}
\usepackage{soul} 
\soulregister\ref{7}
\soulregister\eqref{7}
\soulregister\cite{7}
\soulregister\onlinecite{7}
\usepackage[colorlinks=true, allcolors=blue]{hyperref}

\renewcommand{\st}[1]{}

\newcommand{\be}{\begin{equation}}
\newcommand{\ee}{\end{equation}}

\begin{document}
\title{Exploiting spacetime symmetry in dissipative nonlinear multimode amplifiers for output control  }

\author{Chun-Wei Chen}
\thanks{Equal contributions}
\affiliation{Department of Applied Physics, Yale University, New Haven, CT 06520, USA}

\author{Kabish Wisal}
\thanks{Equal contributions}
\affiliation{Department of Physics, Yale University, New Haven, CT 06520, USA}

\author{Mathias Fink}
\affiliation{Institut Langevin, ESPCI Paris, CNRS, PSL University, 75005 Paris, France}

\author{A. Douglas Stone}
\affiliation{Department of Applied Physics, Yale University, New Haven, CT 06520, USA}

\author{Hui Cao}
\email{hui.cao@yale.edu}
\affiliation{Department of Applied Physics, Yale University, New Haven, CT 06520, USA}

\date{\today}

\begin{abstract}
Time-reversal symmetry enables shaping input waves to control output waves in many linear and nonlinear systems; however energy dissipation violates such symmetry. We consider a saturated multimode fiber amplifier in which light generates heat flow and suffers nonlinear thermo-optical scattering, breaking time-reversal symmetry. We identify a spacetime symmetry which maps the target output back to an input field. This mapping employs phase conjugation, gain and absorption substitution but not time reversal, and holds in steady-state and for slowly varying inputs.  Our results open the possibility of output control of a saturated multimode fiber amplifier.


\end{abstract}


\maketitle



Time-reversal symmetry and reciprocity have been widely explored with various types of waves including electromagnetic, acoustic, and water waves \cite{fink1996time, potton2004reciprocity, przadka2012time}. The inventions of time-reversal mirrors \cite{fink1993time, fink2001acoustic, kuperman1998phase, mounaix2020time} and optical phase conjugators \cite{ya1972correction, bloom1977conjugate, yariv1977amplified} have enabled a wide range of applications, such as aberration correction \cite{nosach1972cancellation, wang1978correction, agarwal1983scattering}, dispersion compensation \cite{yariv1979compensation}, spatial and temporal refocusing \cite{ lerosey2007focusing, yaqoob2008optical, xu2011time, papadopoulos2012focusing, wang2015focusing, horstmeyer2015guidestar, morales2015delivery, dezfooliyan2015spatiotemporal, feldkhun2019focusing, baek2023phase, cheng2023high, bureau2023three}, which have practical significance in imaging, communications, spectroscopy, and sensing~\cite{yariv1978phase, pepper1982nonlinear, rouseff2001underwater, popoff2010image,  fisher2012optical, mosk2012controlling, lerosey2022wavefront, alexandropoulos2022time}.  
In general, time-reversal symmetry holds not only for linear but also nonlinear processes, as long as the waves are not coupled to a bath via an irreversible process \cite{tanter2001breaking, ducrozet2016time, fernandes2022role}. For example, the Kerr-effect-induced self-phase modulation and multi-wave mixing can be reversed to undo pulse distortion \cite{pepper1980compensation}, remove spectral broadening \cite{fisher1983optical}, and even reconstruct rogue waves \cite{chabchoub2014time, ducrozet2020experimental}. 
A common application of time-reversal symmetry is to create a wavefront which can autonomously refocus after scattering by time-reversing the fields generated by a point source in the desired focal position. Once identified, the refocusing wavefront  can be synthesized experimentally and fed back into the same system to achieve focusing. A dissipative process such as linear amplification does break time-reversal invariance, but the input wavefront can still be found by mapping to a different system with loss replacing gain, as has been illustrated e.g., through the study of coherent perfect absorption (time-reversed lasing) \cite{chong2010coherent, wan2011time, pichler2019random, slobodkin2022massively}. In this case, phase conjugation plus replacing gain with loss in the {\it steady-state} wave equation identifies a wavefront that will be completely trapped and absorbed by the time-reversed counterpart. Such mappings which interchange gain and loss in order to find the solution to an inverse problem exist even in the presence of nonlinearity and chaos~\cite{longhi2011time, longhi2012coherent, sweeney2020electromagnetic, suwunnarat2022non}.

The high-power multimode optical fiber amplifier studied here is unlike the nonlinear dissipative systems previously studied; it is characterized by heat diffusion, thermo-optical nonlinearity, and gain saturation~\cite{limpert2002100, cheng2005high, lombard2006beam, jauregui2013high, zervas2014high}. As light undergoes amplification, heat is necessarily generated due to the non-radiative transitions in the pumping cycle, and this heat diffuses irreversibly out of the fiber to the surrounding reservoir. The first derivative of the temperature with respect to time in the thermal diffusion equation reflects this irreversibility and thus breaks time-reversal symmetry. Even if we map this system to a conjugate system with optical gain replaced by an equal amount of absorption, heat will still be generated in the fiber and flow to the reservoir. Thus, the two systems are not time-reversed counterparts in general, and reversing the output signal from a multimode fiber (MMF) amplifier and launching it to the complementary MMF with absorption will not reproduce the original input to the amplifier. Moreover, the non-uniform heating of the fiber, arising from the spatially varying intensity distribution of the multimode interference, causes a non-uniform index change via the thermo-optical nonlinearity and scatters light between fiber modes. As the power increases in the fiber, dynamic mode coupling will destabilize the MMF amplifier and the output beam quality suffers degradation due to an effect termed transverse mode instability (TMI)~\cite{jauregui2020transverse}.

Recent studies reveal that increasing the number of excited modes in a MMF will suppress TMI, enabling further power scaling of fiber amplifiers \cite{chen2023suppressing, wisal2023theoryTMI}. However, even in the absence of TMI, multimode interference will generate speckled fields at output, which is undesired for practical applications. Unlike in a linear MMF without gain, simply phase-conjugating a desired output field profile and sending it back into the fiber amplifier will not generate the required input wavefront. Recently it was shown experimentally that wavefront shaping of a coherent seed to a MMF amplifier can focus the output light~\cite{florentin2017shaping}.  However, it was not known whether it is possible to generate any desired output beam profile of a MMF amplifier by input wavefront shaping.

Here we present a spacetime mapping which shows that indeed an input wavefront of a specific power does exist to generate an arbitrary output beam profile at any chosen power, as long as the fiber amplifier operates below the TMI threshold. Our mapping is a generalization of the more familiar gain--loss mapping between an amplifier and an absorbing counterpart. 
With a monochromatic time-invariant seed, the amplifier reaches a steady state and produces a static output field. In the complementary fiber with absorption, the amplifier output is phase-conjugated and sent back to the distal end. As we will show, the power growth in the nonlinear amplifier with saturated gain is exactly reversed in the absorbing fiber with identical saturation intensity for absorption. Assuming the same amount of heat is generated in the fiber and flows into the thermal bath, the phase-conjugated field undergoes nonlinear thermo-optical scattering that reverses the effect of the nonlinear mode coupling in the amplifier. Therefore, the steady-state transmitted field at the proximal end of the absorbing MMF is identical to the phase-conjugated input of the amplifying fiber, proving that any target amplifier output can be generated by wavefront shaping of a coherent monochromatic seed. 
The spacetime mapping also holds for a dynamic amplifier with time-varying input and output fields. The target dynamic output from the amplifier is phase conjugated but {\it not} time reversed before sending it to the absorbing fiber. The transmitted field provides the phase conjugated input to the amplifier and the resulting output has negligible difference from the desired one. This mapping relies on the fact that the thermal response time ($\sim$ ms) is much longer than the optical response time ($\sim$ ns) in a typical fiber amplifier, and further requires the input/output field envelopes are slowly varying compared to the optical response time, such that dynamic changes of temperature and refractive index are negligible during the time of flight for light through the fiber.  

We now present analytic and numerical results validating this argument. Since the mapping involves connecting mathematically two distinct physical systems, in experiments the optimal input will still need to be found by feedback optimization~\cite{caravaca2013real, florentin2017shaping, gomes2022near, cao2023controlling}. This generalization of symmetry-based mappings beyond time-reversal to described nonlinear dissipative systems not only advances our physical understanding of complex wave phenomena, but also broadens the range of applications for coherent wave control.  



\begin{figure}[hb!]		\centering\includegraphics[width=0.49\textwidth]{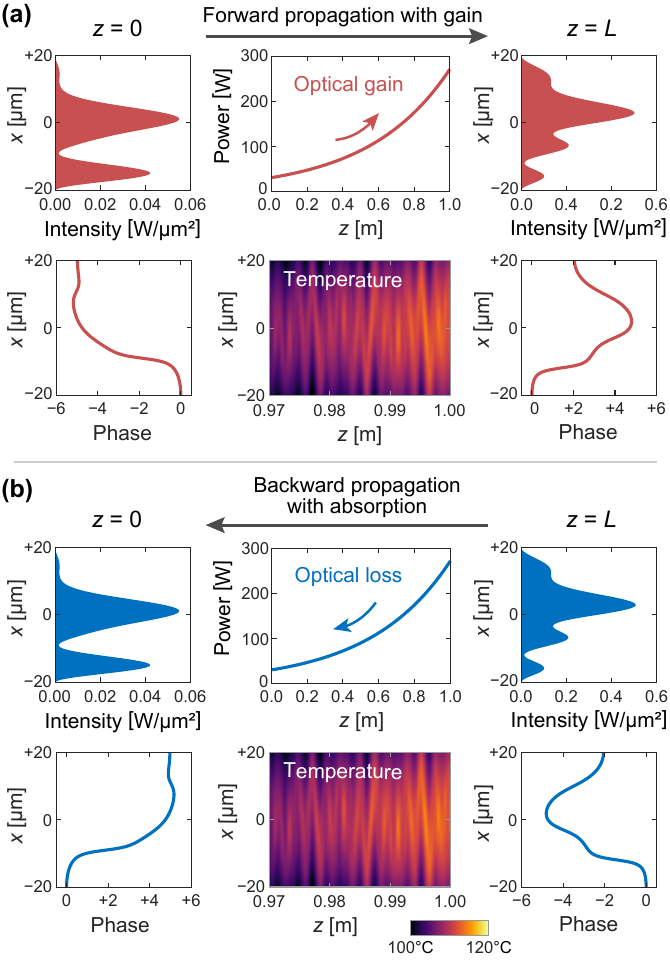}
		\caption{{\bf Steady-state time-reversal of a multimode amplifier with thermo-optical nonlinearity and linear gain.} (a) A 30-W monochromatic input at wavelength 1064 nm with time-invariant wavefront excites five modes in a 1-m-long waveguide amplifier with 1D cross-section (along $x$). With linear, mode-independent gain coefficient $g$ = 2.2 m$^{-1}$, amplified output reaches 270 W. Steady-state temperature profile reveals nonuniform heating caused by inhomogeneous intensity distribution throughout the waveguide. Even without linear mode coupling, thermo-optical coupling and modal dispersion make the output wavefront very different from the input. (b) When the output of the amplifier is phase conjugated and launched backward at the distal end of a complementary waveguide with absorption, the wavefront distortion is undone and the input to the amplifier is recovered (with conjugated phase). The same amount of heat is generated, and the temperature profile is identical to that in (a). In both cases, the temperature at the outer boundary of the waveguide cladding is set to 20$^{\circ}$C. }
        \label{fig:NoGS}
\end{figure}

Consider coherent, narrow-band light of frequency $\omega_0$ launched into a MMF amplifier of length $L$. The electric field can be decomposed by the fiber modes as: 
\begin{equation}
\psi(r_\perp,z,t) = \sum_m  \phi_m(r_\perp) \, A_m(z,t) \, e^{i \beta_m z - i \omega_0 t} \, .
\end{equation} 
The fiber is parallel to the $z$-axis, extending from $z=0$ to $L$, and $r_\perp$ is the transverse coordinate (perpendicular to fiber axis). The $m$-th guided mode in the passive fiber has the transverse field profile $\phi_m(r_\perp)$ and (axial) propagation constant $\beta_m$. $A_m(z,t)$ is the complex amplitude of field in the $m$-th mode at position $z$ and time $t$.  

In the slowly varying approximation for $A_m(z,t)$, the scalar paraxial optical wave equation gives 
\begin{widetext}
\begin{equation}
\frac{\partial A_m(z,t)}{\partial z} + \frac{1}{v_m} \frac{\partial A_m(z,t)}{\partial t} = \left(i\beta_m+\frac{g}{2}\right)A_m(z,t) +
 \sum_j A_j(z,t) \left[ \gamma_{m\,j} + ik_0 \eta e^{i(\beta_j-\beta_m) \, z} \int{\phi_j(r_\perp) \Delta T(r_\perp,z,t) \phi_m(r_\perp) {\rm d} r_\perp} \right], 
\label{Eq:OptFull}
\end{equation}
\end{widetext}
where $v_m$ is the velocity of the $m$-th mode, $g_m$ denotes mode-dependent gain, $\gamma_{m \, j}$ represents linear mode coupling coefficient, and $k_0$ is the vacuum wavenumber.  The frequency of the pump light $\omega_p$ is higher than that of emission $\omega_0$, causing quantum-defect heating $Q(r_\perp,z,t) \propto (\omega_p/ \omega_0 - 1) \, I_{\rm s}(r_\perp,z,t)$ that depends on local intensity $I(r_\perp,z,t) = |\psi(r_\perp,z,t)|^2$.  Multimode interference results in a highly speckled intensity distribution throughout the fiber, and non-uniform heating causes local variation of temperature $\Delta T(r_\perp,z,t)$. Due to thermo-optical nonlinearity, the temperature change induces a refractive-index variation through the thermo-optic coefficient $\eta = 2n({\rm d}n/{\rm d}T)$. The spatial and temporal change of index $\Delta n(r_\perp,z,t)$ introduces nonlinear coupling between fiber modes, represented by the last term of Eq.~\ref{Eq:OptFull}. 
 
The heat diffusion equation is
\begin{multline}
\rho C \frac{\partial \Delta T(r_\perp,z,t)}{\partial t} - \kappa \left(\frac{\partial^2}{\partial r_\perp^2}+\frac{\partial^2}{\partial z^2} \right) \Delta T(r_\perp,z,t) \\ = Q(r_\perp,z,t) = |g(r_\perp,z,t)| \, q_{\rm D} \, I(r_\perp,z,t) \, , 
\label{Eq:HeatFull}
\end{multline}
where $\rho$ is the mass density, $C$ is the specific heat capacity, $\kappa$ is the thermal conductivity, $Q$ is the rate of heat generation per unit volume, $g(r_\perp,z,t)$ is the local gain coefficient, and $q_{\rm D}= \omega_p/ \omega_0 - 1$ represents the quantum defect. 

The thermo-optical nonlinearity involves processes on different time scales (supplemental material, section I~\cite{SM}). The thermal response time, determined by quantum defect heating and thermal diffusion, is on the order of milliseconds in a Yb-doped MMF amplifier. Thus the temperature distribution can be considered static during the transit time ($\sim$ ns) of light in the fiber of typical length $\sim$ 1--10 m. For a given temperature distribution, the optical response time, i.e., the time it takes for optical field distribution throughout the fiber to reach a steady state, is on the order of nanoseconds.


First we consider the case in which the fiber amplifier reaches a steady state when excited with a time-invariant seed signal. In the optical and heat equations, $\partial/\partial t = 0$, and Eqs.~\ref{Eq:OptFull} and~\ref{Eq:HeatFull} are reduced to:
\begin{multline}
\frac{\partial A_m(z)}{\partial z} = \left(i\beta_m+\frac{g_m}{2}\right)A_m(z) + \sum_j A_j(z) \,  \biggl[ \gamma_{m\,j} + 
\\ ik_0 \eta \, e^{i(\beta_j-\beta_m) \, z} \int{\phi_j(r_\perp) \, \Delta T(r_\perp,z) \, \phi_m(r_\perp) \, {\rm d}r_\perp} \, \biggr] 
\label{Eq:OptSS}
\end{multline}
and
\begin{equation}
\begin{aligned}
- \kappa \left(\frac{\partial^2}{\partial r_\perp^2}+\frac{\partial^2}{\partial z^2} \right) \Delta T(r_\perp,z) = |g(r_\perp,z,t)| \, q_{\rm D} \, I(r_\perp,z), 
\end{aligned}
\label{Eq:HeatSS}
\end{equation} 
respectively. 

If $\{A_m\}$ satisfies Eqs.~\ref{Eq:OptSS} and \ref{Eq:HeatSS}, then $\{A^*_m\}$ satisfies the complex conjugate of Eqs.~\ref{Eq:OptSS} and \ref{Eq:HeatSS}, which corresponds to switching gain $g_m$ to absorption $-g_m$ and forward propagation in $z$ to backward propagation $-z$. All coefficients and constants including $q_{\rm D}$ are  kept the same. These equations govern the time-reversed counterpart of the steady-state MMF amplifier. While the field is complex-conjugated, the intensity distribution is unchanged. While the atomic transitions in the absorbing fiber will not be reversed, we simply assume that the same amount of heat will be generated by absorption in the fiber. Then the temperature distribution will be unchanged, as the thermal boundary conditions are identical. Since the equations for the time-reversed counterpart of the steady-state MMF amplifier are invariant, ``sending back" the conjugated output will produce the original input field, conjugated.   


\begin{figure*}[hbt!]		\centering\includegraphics[width=0.95\textwidth]{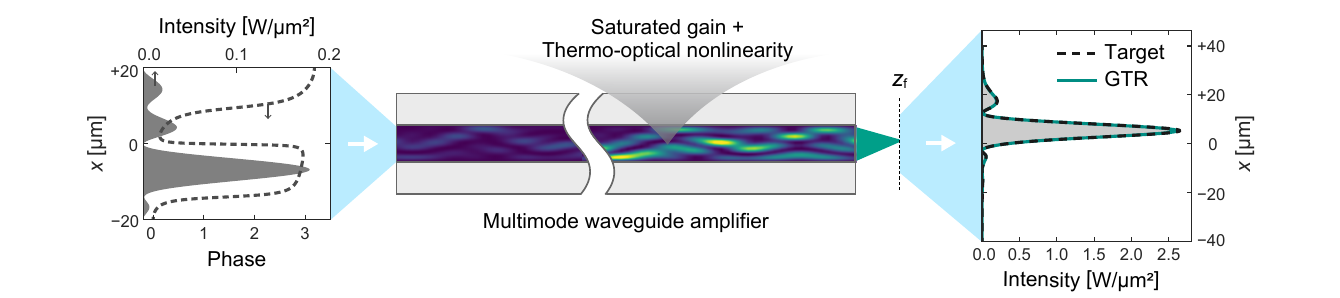}
		\caption{{\bf Focusing through a nonlinear multimode amplifier with gain saturation.} Shaping the input wavefront of a monochromatic seed (at 60 W) to a five-mode waveguide amplifier with thermo-optical nonlinearity and gain saturation to focus the amplified output (at 750 W) to a designated location ($z_{\rm f}$ = 116 \textmu m beyond the end facet and 5 \textmu m from the waveguide axis). The small-signal gain coefficient is 7.0 m$^{-1}$, and the saturation intensity is 0.0625 W/\textmu m$^2$. The input wavefront is obtained by sending the focused output through a complementary waveguide with saturable absorption and then phase-conjugating the transmitted field profile. The focused output from the amplifier exhibits a diffraction-limited spot of width 10 \textmu m, and small side lobes can be significantly suppressed by increasing the number of waveguide modes that contribute to output focusing. }
        \label{fig:GainSat}
\end{figure*}

To confirm this argument for the steady-state condition, we perform numerical simulations in the time domain. We consider a waveguide of 1D cross section with core width $w$ = 40 \textmu m, cladding width $W$ = 400 \textmu m, refractive index is $n$ = 1.5, and length $L$ = 1 m. For simplicity, linear mode coupling is neglected, $\gamma_{m\,j} = 0$, and we assume the optical gain $g_m$ is linear (gain saturation is considered later) and identical for all modes. Additional details of the simulation and relevant time scales are given in ~\cite{SM}.


Figure~\ref{fig:NoGS} shows the numerical result for one representative example. A coherent, monochromatic seed is launched into the waveguide at $z = 0$ and amplified from 30 W at $z=0$ to 270 W at $z=L$. The steady-state output field profile is very different from the input, due to modal dispersion and thermo-optical mode coupling. The output field pattern is phase-conjugated and launched into a waveguide with absorption ($g_m \to -g_m$) from the distal end. The power is reduced back to 30 W at the proximal end, and the steady-state output field pattern of the absorptive waveguide is indeed the same as the original input to the amplifier (with conjugated phase) to high accuracy, while the temperature distribution is identical in the two cases, validating our argument for the steady-state case.

Next, we include the additional saturating nonlinearity in the multimode amplifier. At steady state, $g_m = g_m^{(0)}/[1+I(r_\perp,z)/I_{\rm sat}]$, where $g_m^{(0)}$ denotes the small-signal gain for the $m$-th mode of the fiber, and $I_{\rm sat}$ is the saturation intensity of a MMF amplifier. The presence of gain saturation in real amplifiers has important physical effects. Since the degree of saturation depends on the spatially varying intensity, the gain distribution becomes spatially inhomogeneous in both the transverse 
and longitudinal directions. Hence the nonlinear gain becomes mode-dependent, modifying the growth rate of individual modes and their interference throughout the fiber. The resulting intensity changes alter the heat generation and temperature distribution. Consequently, the thermo-optical coupling between fiber modes is significantly modified by gain saturation, and the output field pattern changes dramatically.  

However the time-reversed counterpart of a steady-state MMF amplifier still exists, with saturated gain replaced by saturated absorption. In the time-reversed counterpart, optical absorption $-g_m$ is saturated by the same amount as optical gain, because the intensity distribution is unchanged and $I_{\rm sat}$ is identical. We verify this explicitly with time-domain numerical simulations. The results are shown in Fig.~\ref{fig:GainSat} (details in the caption). Here the time-reversed counterpart is utilized to obtain an input wavefront that focuses to a diffraction limited spot after propagating through the waveguide amplifier with gain saturation and thermo-optical nonlinearity. This opens up important practical applications in obtaining high-quality output beams using multimode fiber amplifiers, which can provide much higher stable output power than single-mode fiber amplifiers~\cite{chen2023suppressing}.

 \begin{figure*}[thb!]		\centering\includegraphics[width=0.95\textwidth]{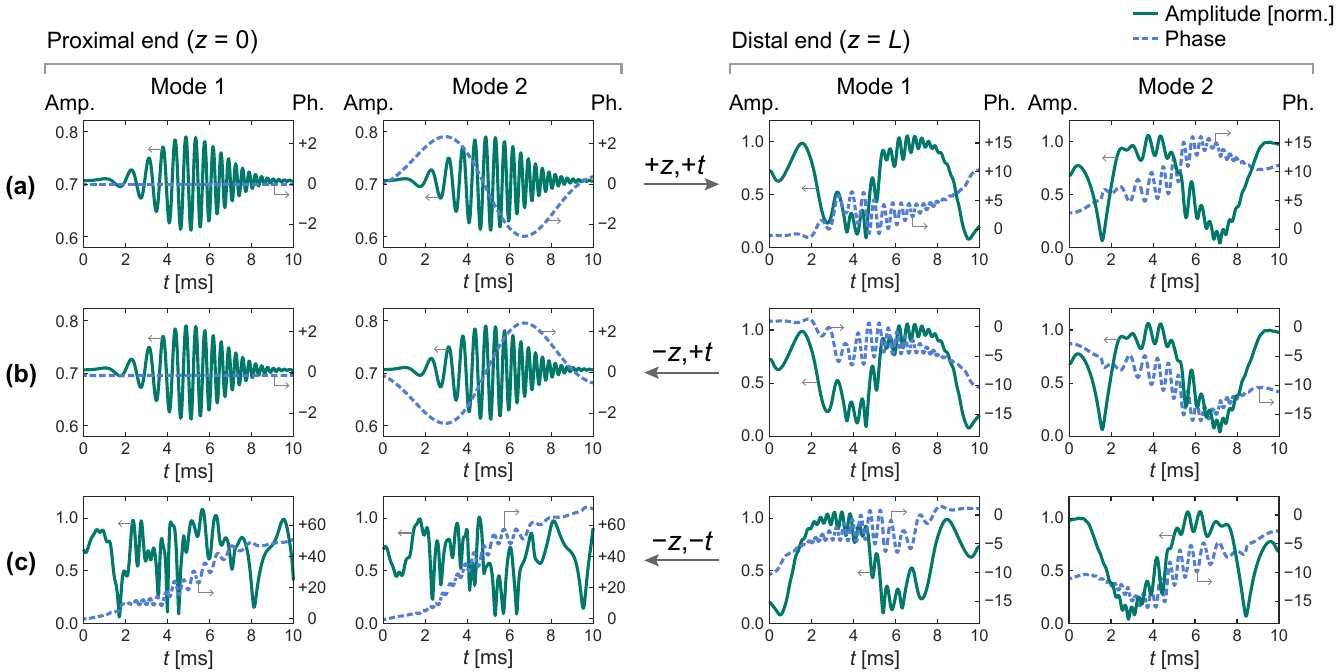}
		\caption{{\bf Generalized time reversal of a dynamic nonlinear amplifier.} Time traces of field amplitude (amp.) and phase (ph.) of a two-mode waveguide amplifier at proximal and distal ends. (a) A time-varying field with a peak power of 237 W is launched into a 0.1-m-long waveguide amplifier with two modes equally excited at $z=0$. Input field amplitude of both modes has the same Gaussian envelop and is temporally chirped at $\sim$2.1 kHz. While mode 1 has a time-invariant phase, mode 2 has a phase modulation at $\sim$0.1 kHz. Amplified field reaches a peak power of 300 W at $z=L$. Heat-induced dynamic mode coupling in the waveguide causes spatio-temporal distortion of output field. In each diagram, the time-varying mode amplitudes are normalized (norm.) by the time-averaged total power. (b) Phase-conjugation without time reversal of the amplifier output is sent to the absorbing waveguide, and the transmitted field is almost identical to the phase-conjugated input to the amplifier. (c) Phase-conjugated, time-reversed output field of the amplifier is launched to a complementary waveguide with absorption at $z=L$, and the transmitted field at $z=0$ is totally different from the amplifier input. }
        \label{fig:TimeVarying}
\end{figure*}


 In the previous examples the multimode amplifier reaches a steady state with a time-invariant seed, where the time derivative of the envelope drops out.  We now consider a dynamic state of the amplifier, in which both the amplitude and phase of the input envelope vary on the time scale of 0.1--1 ms, similar to that of the temperature changes.   In Fig.~\ref{fig:TimeVarying}a we show an example of the output generated by the amplifier for such a seed.  For computational simplicity, in this time-varying case we consider only two modes in a 0.1-m-long fiber and neglect gain saturation.  Analyzing Eqs.~\ref{Eq:OptFull} and~\ref{Eq:HeatFull}, we see that phase conjugating the output field and sending it back in the absorptive waveguide {\it without} reversing the envelope in time  corresponds to the mapping $z \rightarrow -z$, $t \rightarrow t$, $g_m \rightarrow -g_m$, $A_m(z,t) \rightarrow A_m^*(-z,t)$, and $\Delta T(r_\perp,z,t) \rightarrow \Delta T(r_\perp,-z,t)$. The heat equation Eq.~\ref{Eq:HeatFull} remains invariant, but the optical equation Eq.~\ref{Eq:OptFull} is not due to the $ (1/v_m) \partial A_m(z,t)/\partial t$ term. However, this term is much smaller than $ \partial A_m(z,t)/\partial z$ and can be neglected, as the optical pulse envelope evolves temporally on the thermal response time which is much longer than the nonlinear optical response time in the fiber. Ignoring $ (1/v_m) \partial A_m(z,t)/\partial t$ makes the optical equation also remain invariant under our proposed mapping.
Hence we propagate backwards in space the conjugated output of Fig.~\ref{fig:TimeVarying}a in Fig.~\ref{fig:TimeVarying}b without reversing the direction of time, and indeed find that it generates a transmitted field almost identical to the phase-conjugated input to the waveguide amplifier. We note that the naive time-reversal operation with additional mapping $t \to -t$
satisfies the equation for light propagation Eq.~\ref{Eq:OptFull}, but not that for heat diffusion Eq.~\ref{Eq:HeatFull}. Hence this mapping does not reproduce the conjugated input field, as confirmed numerically in Fig.~\ref{fig:TimeVarying}c. A phase-conjugated, time-reversed output field propagated backward in the waveguide with absorption, creates a transmitted field completely different from the original input to the waveguide amplifier [Fig.~\ref{fig:TimeVarying}c]. 






In summary, we have identified a spacetime symmetry mapping which demonstrates the existence of an input field profile at a specific power which generates a desired target output field of a chosen power for a  MMF amplifier with both thermo-optical and saturating nonlinearities.  This mapping is valid for both steady-state inputs and slowly varying pulses, but will fail when the power is high enough to cause a dynamical instability, such as TMI (see further discussion in \cite{SM}).  This limitation is similar to other nonlinear systems in the regime of dynamical chaos or instability, where even if time-reversal symmetry exists, noise and sensitivity to small perturbations make it impractical to exploit.  
  

Our results extend the understanding of the possibilities of wave control after nonlinear propagation in dissipative media.  In particular it implies the possibility of generating any output beam profile of a high power amplifier, including a diffraction-limited focal spot that can be collimated subsequently to the far field. Given the concerns of output beam quality for high power fiber amplifiers~\cite{limpert2002100, cheng2005high, lombard2006beam, jauregui2013high, zervas2014high}, our results pave the way for employing multimode fiber amplifiers in high-power applications, leveraging their advantages over the single-mode counterparts, e.g., high power thresholds for transverse mode instability~\cite{chen2023suppressing, wisal2023theoryTMI} and stimulated Brillouin scattering~\cite{chen2023mitigating, wisal2023theory}. 
We note that such wavefront-shaping schemes work well only for narrow-band amplifiers; specifically, when the spectral bandwidth of an input signal is less than the spectral correlation width of the output field pattern. 


We thank Yaniv Eliezer for assisting in establishing the time-domain simulation code. We acknowledge Alexey Yamilov, Owen D.~Miller, and Shanhui Fan for fruitful discussions. This work is supported by the Air Force Office of Scientific Research (AFOSR) under Grant FA9550-20-1-0129 and by the Simons Foundation. We acknowledge the computational resources provided by the Yale High Performance Computing Cluster (Yale HPC).


\begin{appendices}
\appendix

\section{Optical and thermal time scales}
As noted in the main text, the spacetime mapping which is the subject of this work is approximately valid due to different times scales involved in the 
thermo-optical nonlinearity.  The physically relevant processes involve both optical and thermal time scales, which can differ by several orders of magnitude. In this section, we define and provide an estimate for various timescales mentioned in the main text.   

\vspace{-5mm}
\subsection{Thermal response time}
\vspace{-2mm}

The thermal response time is defined as the time it takes for the fiber toreach a steady-state temperature distribution for a given optical intensity distribution in a multimode fiber (MMF). It includes both the heating time and thermal diffusion time. 

Quantum defect heating occurs on the timescale dictated by transition rates between energy levels in the gain medium. For Yb-doped silica fibers, this heating is on the order of 1 ms. 

Nonuniform heating from optical speckles formed by multimode interference results in an inhomogeneous temperature distribution in the fiber. The thermal gradients induce thermal diffusion in both the transverse and longitudinal directions in the fiber. Assuming a fixed temperature at the cladding outer boundary, the temperature gradient in the transverse direction is much larger than that in the longitudinal direction. Thus heat diffusion occurs predominantly in the transverse direction. The characteristic time scale of heat diffusion over the fiber cross-section is estimated as $\tau_{\rm D} \simeq {R^2}/{D}$, where $D$ is the diffusion coefficient and $R$ is the fiber core radius. For a typical silica multimode fiber of $R=20$ \textmu m and $D\approx 10^{-6}$ $\rm m^2/s$, the diffusion time $\tau_{\rm D} \approx 0.4$ ms. 

\vspace{-5mm}
\subsection{Optical response time}
\vspace{-2mm}

The optical response time is defined as the time it takes to reach a steady-state light distribution in a MMF for a given temperature distribution. This response involves both the thermo-optical process that causes a refractive index change and the resulting nonlinear mode
coupling in the fiber. The nonlinear coupling length $L_{\rm NL}$ is defined as the length scale over which significant nonlinear mode coupling occurs. Its value is given by $L_{\rm NL}^{-1}= g\chi_0 P_{\rm s}$ where $g\: [{\rm m}^{-1}]$ is the linear optical gain coefficient, $\chi_0$ is the nonlinear thermo-optic coupling coefficient with units $[{\rm W}^{-1}]$, and $P_{\rm s}\:[{\rm W}]$ denotes the signal power. A typical operating condition of the multimode amplifiers discussed in the main text would give, $g \sim 1$ m$^{-1}$, $\chi_0 \sim 0.1$ $ {\rm W}^{-1}$, and $P_{\rm s} \sim 100$ W, we estimate $L_{\rm NL}$ is of the order 0.1 m. The nonlinear optical response time $\tau_{\rm NL}$ is defined as the propagation time of light over the nonlinear coupling length $L_{\rm NL}$.  For $L_{\rm NL} \sim 0.1$ m, $\tau_{\rm NL}$ is roughly 0.5 ns.

\vspace{-5mm}
\subsection{Seed modulation time}
\vspace{-2mm}

Since the diffusion time $\tau_D$ is on the order of millisecond, this results in maximal nonlinear thermo-optic coupling at frequencies of the order of kHz. Therefore the non-linear thermo-optic interaction will mainly affect the envelope of the coherent seed when it is modulated on the timescale of 0.1--1 ms ($\sim \tau_{\rm D}$), which is much longer than the non-linear optical response time (0.5 ns) defined above. 

\vspace{-5mm}
\subsection{Consequence of timescale mismatch}
\vspace{-2mm}

The above estimates reveal that thermal and optical response times differ by roughly six orders of magnitude; hence light propagation over the nonlinear coupling length is practically instantaneous compared to temporal evolution of the optical envelope. In the light propagation equation (Eq.~2 in the main text), the term with longitudinal derivative of the mode envelope has the magnitude $|v_{\rm m}{\partial A_m}/{\partial z}| \approx |v_m{A_m}/{L_{\rm NL}}| \approx |{A_m}/{\tau_{\rm NL}}|$, where $A_m$ is the optical amplitude of $m$-th fiber mode, $v_m$ is its effective velocity. In contrast, the relevant modulation time of an input signal is on the order of thermal response time ($\sim \tau_{\rm D}$), for which the term with time derivative of $A_m(z,t)$ in optical wave equation has the magnitude $|{\partial A_m}/{\partial t}| \approx |{A_m}/{\tau_{\rm D}}|$. The ratio of these two terms is  
\begin{align*}
\frac{|{\partial  A_m}/{\partial t}|}{|v_m {\partial  A_m}/{\partial z}|} \approx \frac{\tau_{\rm NL}}{\tau_{\rm D}} \approx 10^{-6}.
\end{align*}
Thus, as noted in the text, the temporal derivative term in the optical wave equation (Eq.~2 in the main text) can be neglected. It is for this reason that when the phase conjugated wavefront is sent back to an absorptive fiber {\it without} reversing the temporal envelope, the transmitted field matches the phase-conjugated input to the fiber with equivalent gain to very high accuracy, and our spacetime transformation can be applied beyond the steady-state condition. 

\begin{figure*}[hbt!]		\centering\includegraphics[width=0.95\textwidth]{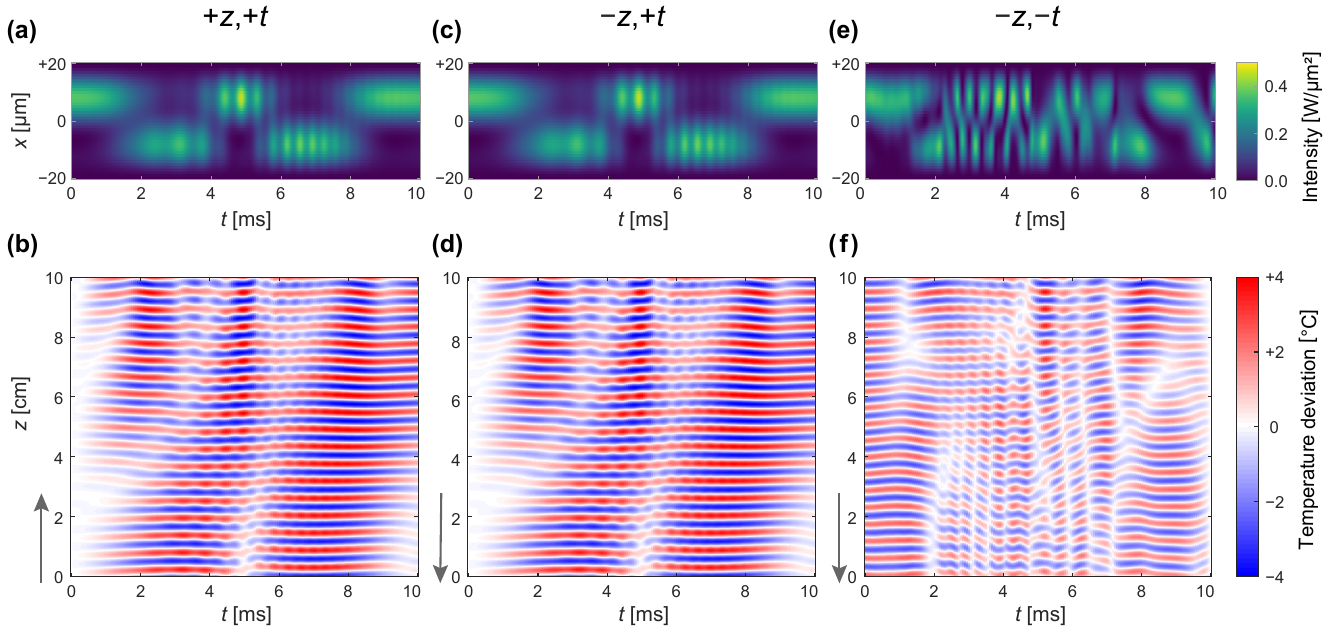}
	\renewcommand{\thefigure}{S\arabic{figure}}
	\caption{{\bf Spatio-temporal variations of light intensity and temperature in a dynamic amplifier and its counterpart.} A time-varying seed (of field intensity in a) is launched to a two-mode waveguide with linear gain and thermo-optical nonlinearity. It generates a moving temperature grating (shown at $x$ = +10 \textmu m in b), causing dynamic mode coupling. (c,d) The amplifier output is phase-conjugated but \textit{not} time-reversed, then launched into the distal end of a waveguide with equivalent absorption $(-z, +t)$. Transmitted field intensity (c) reproduces that of original input to the amplifier (a), and the temperature distribution (d) is also recovered. (e,f) The amplifier output is phase-conjugated \textit{and} time-reversed, then propagates backward in the absorptive waveguide $(-z, -t)$. The transmitted field intensity at the proximal end (e) is distinct from the original seed to the amplifier (a), and the temperature distribution (f) also differs from that in the amplifier (b). }
	\label{fig:TimeVarying}
\end{figure*}

\section{Numerical simulation}

The simulation parameters are chosen to be comparable to those of a ytterbium (Yb)-doped silica fiber~\cite{hansen2013theoretical}. The waveguide is parallel to the $z$ axis, and has a one-dimensional cross-section along the $x$ axis. The core is 40 \textmu m wide and surrounded by a cladding of width 400 \textmu m. The refractive index of the core is 1.5 at 20$^{\circ}$C, and rises with increasing temperature according to the thermo-optic coefficient $\eta = 2n({\rm d}n/{\rm d}T) = 3.51 \times 10^{-5}$ K$^{-1}$. For both core and cladding, the thermal conductivity $\kappa = 1.4\times 10^{-6}$ WK$^{-1}$\textmu m$^{-1}$, and the product of mass density and specific heat capacity $\rho C = 1.67\times 10^{-12}$ JK$^{-1}$\textmu m$^{-3}$, are typical values of silica. The outer boundaries of the cladding are thermally conducting (the Dirichlet boundary condition) and kept at 20$^{\circ}$C. For simulations with linear gain [Figs.~1 and 3], the gain coefficient $g_m = 2.2$ m$^{-1}$ is independent of power and mode index. For simulations considering gain saturation [Fig.~2], the gain coefficient depends on the local intensity $I(x,z,t)$ as $g = g^{(0)}/[1+I(x,z,t)/I_{\rm sat}]$, in which $g^{(0)} = 7.0$ m$^{-1}$ and $I_{\rm sat} = 0.0625$ W\textmu m$^{-2}$. Pump depletion is neglected.  

Our code solves the coupled optical and thermal equations (Eqs.~2 and 3 in the main text) iteratively using finite-difference methods. The spatial and temporal discretizations are carefully chosen to achieve sufficient numerical accuracy and convergence. When the input signal is time-invariant, the thermal response time is much slower than its optical counterpart, and the temperature distribution can be treated as static during the time that light travels through the waveguide. For a given temperature distribution at the time step $t_N$, light propagation in the waveguide is solved with Eq.~2 in the modal basis with a longitudinal-grid size of $\sim$0.1--1 \textmu m. The resulting mode amplitude $A_m(z, t_N)$ throughout the waveguide is converted to a spatial intensity distribution $I(x,z,t_N)$ in order to update the heat generation rate $Q(x,z,t_N)$ in the thermal equation (Eq.~3). Using the Crank--Nicolson method, the heat diffusion equation is solved in real space with grid size $\Delta z \sim$ 100 \textmu m and $\Delta x \sim$ 0.1 \textmu m to obtain the temperature distribution $T(x,z,t_{N+1})$ at the next time step $t_{N+1}=t_N+\Delta t$. In the simulation, we neglect longitudinal heat diffusion, because the temperature gradient in $z$ is much smaller than that in $x$. For a given static input, this process reiterates with a time step $\Delta t = 100$ ns for about $10^6$--$10^7$ times to reach a steady state.   

\section{Temporal modulation of seed}

For simulations with dynamic input signals, higher numerical accuracy is needed in simulating thermo-optical effects, and we reduce both the spatial grid size and temporal step size in the simulation. To avoid extremely long simulation time, we shorten the waveguide to 10 cm, and consider only two guided modes without gain saturation. The input signal is modulated over a constant background. We first simulate the amplifier with only the constant background as input. The amplifier eventually reaches a steady state with equal mode excitation at the input power 190 W. With a gain coefficient $g$ = 2.2 m$^{-1}$, the power output reaches 237 W at the steady state. 

The temporal modulation of the input signal has a Gaussian envelope, as shown in Fig.~3a of main text. The input field amplitudes in both waveguide modes are chirped in time. The center frequency of modulation is $\sim$2.1 kHz. The maximum peak-to-peak modulation is $\sim$100 W. While the input field in the fundamental mode (mode 1) has a time-invariant phase, the high-order-mode (mode 2) has an input phase modulated at $\sim$0.1 kHz. The resulting spatio-temporal intensity profile at the input is displayed in Fig.~\ref{fig:TimeVarying}a. This time-varying field propagates forward in the waveguide ($+z$), resulting in dynamic coupling between the two modes via the thermo-optical nonlinearity in the amplifier. In Fig.~\ref{fig:TimeVarying}b, we plot the temperature deviation from its steady-state value at one transverse position $x$ = +10 \textmu m throughout the waveguide to illustrate thermal fluctuations. The thermo-optical interaction can be seen from the time-varying field that leads to formation of a moving temperature grating. This dynamic mode coupling generates a time-varying output field which is recorded for subsequent backward-propagation ($-z$) simulations. 

\begin{figure*}[t!]		\centering\includegraphics[width=0.95\textwidth]{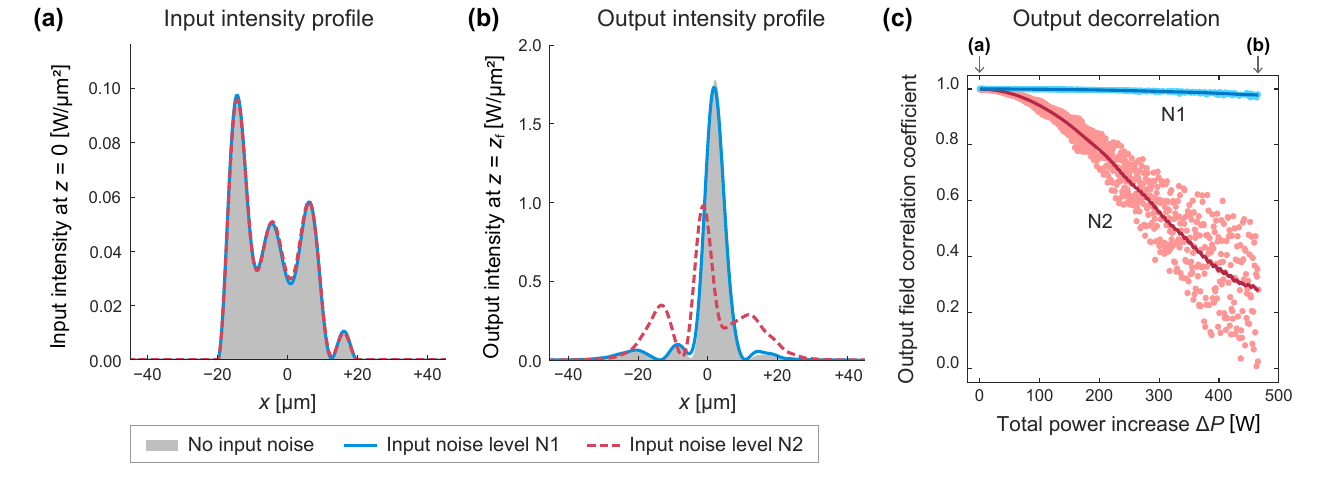}
	\renewcommand{\thefigure}{S\arabic{figure}}
	\caption{{\bf Effect of input amplitude noise on output field profile.} A five-mode waveguide amplifier is simulated in the time domain. Applying the spacetime symmetry mapping, a time-invariant input field profile of 60 W at $z=0$ is obtained for focusing output light of 525 W at 180 \textmu m away from the output facet ($z=z_{\rm f}$) to a diffraction-limited spot. Then random amplitude noise is introduced to the input wavefront, with the standard deviation of the spatial amplitude variations equal to 0.004 (0.014) of the mean amplitude for the noise level N1 (N2). (a) Transverse intensity profiles of the input intensity distribution (at $z=0$) without noise (gray), with noise N1 (blue solid) and N2 (red dashed). The amplitude noise has negligible effect on the input wavefront, but it grows along the fiber due to optical amplification and thermo-optical nonlinearity. (b) Transverse intensity profiles at the target focal plane outside of the amplifier output ($z=z_{\rm f}$), showing that the output beam remains focused for N1, but completely distorted for N2. (c) Pearson correlation between the generated and target field profiles decreases with power increase from amplifier input to output $\Delta P=P(z=L)-P(z=0)$. Curve: mean correlation as a function of $\Delta P$. At small $\Delta P$, the field correlation is almost 1, meaning that the generated field perfectly reproduces the target wavefront. The correlation drops with increasing $\Delta P$, and the rate is higher for larger input noise. }
	\label{fig:Noise}
\end{figure*}

\section{Generalized time reversal}

To confirm that this multimode amplifier is not invariant under time-reversal, we reverse the temporal variation of the recorded output field and generate a phase-conjugated replica. Subsequently, this time-reversed signal is launched into the distal end of the waveguide with equivalent absorption to the amplifier gain, and propagates backward to the proximal end $(-z, -t)$. The numerical simulation starts with a temperature distribution matching that of the amplifier recorded after the seed modulation ends. The transmitted field profile at the proximal waveguide end is recorded and shown in Fig.~\ref{fig:TimeVarying}e. {\it Its evolution is markedly different} from the modulated input to the waveguide amplifier [Fig.~\ref{fig:TimeVarying}a]. Figure~\ref{fig:TimeVarying}f reveals that the temperature distribution at $x$ = 10 \textmu m also differs dramatically from that recorded with forward-propagation in the amplifier. We recall the origin of the failure to generate a time-reversed replica output by revisiting the coupled optical and heat equations (Eqs.~ 2 and 3 in the main text). While reversing both $z$ and $t$ with phase conjugation preserves the optical equation (Eq.~2), it does not leave the heat equation (Eq.~3) invariant due to the presence of a first-order time derivative of the temperature fluctuation, namely, the term $\partial \Delta T / \partial t$ changes the sign with time reversal $t \rightarrow -t$. Although the optical equation also has a term containing the first-order derivative in time, Eq.~2 remains invariant due to the combination of complex conjugation {\it and} reversing the time $t \rightarrow -t$. 

As noted in the text,in our spacetime transformation we keep the heat equation invariant by not reversing temporal envelope of amplifier output. Hence in our simulations we follow the backward propagation of phase-conjugated output, while keeping the temporal envelope the same $(-z, +t)$. The initial temperature distribution in the absorptive waveguide is the same as the initial temperature distribution in the amplifier. As shown in Figs.~\ref{fig:TimeVarying}c and~\ref{fig:TimeVarying}d, the spatio-temporal fluctuation of temperature in the absorbing waveguide is almost identical to that in the amplifying one, and the proximal-end field profile reproduces the phase-conjugated seed and its evolution in time. Without reversing the temporal envelope of output signal from the amplifier, now the time derivative term in the optical wave equation (Eq.~2 in the main text) does not transform correctly, but as noted above, this term is effectively negligible, Therefore, our scheme allows retrieval of a time-varying wave launched at the distal end of a nonlinear dissipative system.  

\section{Instability at high power}
In theory, this spacetime symmetry transformation should be applicable to a multimode fiber amplifier at any power level. However, the presence of seed noise, pump noise, and/or external perturbations to the amplifier can potentially disrupt the spacetime symmetry mapping, particularly at high power. This is because such noise will be amplified along the fiber not only because of optical gain, but also by the thermo-optical nonlinearity. As an example, we show numerically how adding a small amplitude noise to the input seed can cause the output field to deviate from the target profile in Figs.~\ref{fig:Noise}a and~\ref{fig:Noise}b. Here we consider a five-mode amplifier with linear gain of $\sim$2.2 m$^{-1}$ and target a focused beam of 525 W at 180 \textmu m away from the output facet. We first send the phase conjugate of the steady-state output field through a complementary absorptive waveguide to obtain the input wavefront $\psi_{\rm 0}(x,z=0)$ at the power of 60 W for the amplifier. In the absence of noise, launching such a seed into the amplifier can produce the original output profile when reaching the steady state (gray areas in Figs.~\ref{fig:Noise}a and~\ref{fig:Noise}b).
Since we are considering steady-state, we perturb the correct input with static random amplitude noise in each mode. The standard deviation of the spatial amplitude variations, $\sigma_x(|\psi_{\rm N}(x,z=0))|-|\psi_{\rm 0}(x,z=0)|)$, is 0.004 and 0.014 of the mean amplitude, $\langle|\psi_{\rm 0}(x,z=0))|\rangle_x$, for two noise levels of N1 and N2, respectively. The perturbations are so small at the input that the wavefront $\psi_{\rm N}(x,z=0)$ appears almost identical to the original seed [Fig.~\ref{fig:Noise}a]. However, as the signal power is increased during propagation in the gain fiber, the noise is also amplified. With the smaller input noise N1, the output beam profile at the target focal plane $\psi_{\rm N}(x,z=z_{\rm f})$ only slightly deviates from the target profile, and still remains focused (blue solid curve in Fig.~\ref{fig:Noise}b). However when the input noise is increased to N2, the beam profile of the generated output (red dashed curve in Fig.~\ref{fig:Noise}b) becomes highly distorted from the target $\psi_{\rm 0}(x,z=z_{\rm f})$. 

Since the thermo-optical nonlinearity is power-dependent, the resulting noise amplification should increase with the signal power in the amplifier. The total power increase through the fiber of length $L$ is $\Delta P = P(z=L)-P(z=0)$. We compute the Pearson correlation between $\psi_{\rm N}(x,z=z_{\rm f})$ and $\psi_{0}(x,z=z_{\rm f})$ at different $\Delta P$, which effectively quantifies the deviation of the generated wavefront from the target. Figure~\ref{fig:Noise}c shows that, when $\Delta P$ is small, the field correlation coefficient is close to unity, indicating that the generated wavefront at the amplifier output successfully replicates the target profile. This coefficient drops more rapidly as $\Delta P$ increases. Figure~\ref{fig:Noise}c also reveals that the larger the input noise, the higher rate of the output decorrelation. 

While here we have considered the effect of steady-state perturbations, we expect the dynamic spacetime symmetry mapping to break down at high power due to the dynamic instability seeded by noise, such as the transverse mode instability (TMI) in a high-power fiber laser amplifier~\cite{jauregui2020transverse}. In this regard, it is worth noting that elsewhere we showed that TMI can be suppressed by simultaneously exciting many modes coherently in a multimode fiber amplifier~\cite{chen2023suppressing}; where such a multimode excitation can be easily realized by, e.g., focusing the output light to a diffraction-limited spot in the near field of the fiber distal end.  
\end{appendices}

\bibliography{myref}

\end{document}